\documentstyle[aps,prl,floats,epsf,epsfig]{revtex}
\tighten
\draft

\begin{document}


\wideabs{

\title{Nuclear Propelled Vessels and Neutrino Oscillation Experiments}

\author{J.~Detwiler, G.~Gratta, N.~Tolich, Y.~Uchida}
\address{Physics Department, Stanford University, Stanford, CA}

\date{\today}

\maketitle

\begin{abstract}

We study the effect of naval nuclear reactors on the study of neutrino oscillations.
We find that the presence of naval reactors at unknown locations and times may limit
the accuracy of future very long baseline reactor-based neutrino oscillation experiments.
At the same time we argue that a nuclear powered surface ship such
as a large Russian ice-breaker may provide an ideal source for precision
experiments.  While the relatively low reactor power would in this case require
a larger detector, the source could be conveniently located at essentially any distance
from a detector built at an underground location near a shore in a region of the world 
far away from other nuclear installations.  The variable baseline would allow for a precise 
measurement of backgrounds and greatly reduced systematics from reactor flux and detector 
efficiency.   In addition, once the oscillation measurement is completed, the detector
could perform geological neutrino and astrophysical measurements with minimal reactor 
background.

\end{abstract}


}


Reactor-based neutrino oscillation experiments~\cite{Bemporad_etal} have recently generated renewed 
interest because of their ability to probe very small $\Delta m^2$ values albeit with modest 
$\sin^2 2\theta$ sensitivity.
The $\sim 1$~km baseline experiments at the Chooz~\cite{Chooz} and Palo Verde~\cite{PaloVerde} 
reactors were optimized for the $10^{-3}~{\rm eV^2}\lesssim \Delta m^2 \lesssim 10^{-2}~{\rm eV^2}$
region and the large mixing angle suggested by the atmospheric neutrino anomaly~\cite{atm_nu}.  
The negative results from these two experiments contributed to the elimination of $\nu_{\rm e}$ as 
a main component of the atmospheric neutrino signal.    The KamLAND experiment~\cite{KamLAND}, now 
taking data, was designed with a $\approx 200$~km baseline making it
sensitive to the oscillation parameters in the large mixing angle solution (LMA-MSW) to the solar 
neutrino problem~\cite{solar_nu}.  In addition the Borexino detector~\cite{borexino}, although 
designed to detect the $^7$Be neutrinos from the sun, might also perform oscillation measurements 
using European reactors~\cite{schoenert}.  


In the future the interest for reactor-based neutrino oscillation experiments may shift towards
the possibility of performing precision measurements.   A proposal has been put forward~\cite{Mikaelyan}
to improve the measurement of $\sin^2 2\theta_{13}$ beyond the $\sim 0.1$ sensitivity of Chooz and 
Palo Verde with a high statistics, $\sim$1~km baseline (double) detector.
If the MSW solution is confirmed, depending on the $\Delta m^2$ value found by KamLAND, it may be
appropriate to build either a larger detector ($\sim$10-20~kton) with a similar baseline~\cite{svoboda}
or a somewhat smaller detector with a $\sim 10-20$~km baseline\cite{Petcov,schoenert-lasserre} to perform 
precision measurements.   

Very little attention has been given until now to the background due to nuclear propelled vessels to 
these experiments.   Although we estimate that the total thermal power generated by such vessels is 
just a few percent of the total nuclear thermal power in the world, we will show that the mobility and stealthy 
nature of these installations may produce backgrounds for the present generation experiments.
In the case of future detectors, awareness of the background from naval reactors and careful geographical 
placement will be essential to retain the ability to perform meaningful measurements.

It is also interesting to consider whether this apparent liability could be turned into
an asset, by placing the detector far away from any fixed nuclear installations and bringing in a reactor
on a vessel.  The variable baseline available this way would be important to cancel most systematics 
and tune the experiment to some particular set of oscillation parameters.     In addition, in the case 
of a very large detector, removing the reactor altogether would allow for lower backgrounds in the 
study of geological~\cite{terr} and astrophysical neutrinos.


\begin{table*}[t!!!]
\begin{tabular}{|l|c|c|c|c|c|c|}
  Case                                   & Power       & Det. Mass & Distance   & Flux at 100\% power & Rate        & Fraction  \\
                                         & (GW$_{th}$) & (kton)    & (km)       & (cm$^{-2}$s$^{-1}$) & (yr$^{-1}$) & of Signal \\
\hline
KamLAND~\cite{KamLAND}                   &   175.7     &    1      & $\sim 200$ & $1.3\times 10^6$    & 880         &     ---   \\
``Typhoon'' sub in Toyama Bay            &    0.38     &    1      & 40         & $1.8\times 10^5$    & 92          &   0.10    \\
Aircraft carrier in Yokosuka             &    0.84     &    1      & 200        & $9.5\times 10^3$    & 8.1         &   0.01    \\
\hline
Borexino~\cite{schoenert}                & $\sim 520$  &    0.3    & $\sim 800$ & $1.9\times 10^5$    & 27          &     ---   \\
``Typhoon'' sub near Giulianova$^{\ast}$ &    0.38     &    0.3    & 50         & $6.9\times 10^4$    & 12          &   0.46    \\
Aircraft carrier in Naples               &    0.84     &    0.3    & 200        & $9.5\times 10^3$    & 1.7         &   0.06    \\
\end{tabular}
\caption{Comparison between fixed reactor signal and potential background from naval reactors for KamLAND and
Borexino.    Since these experiments derive measure neutrinos from a large set of reactors, the distance and power 
figures for fixed reactors (lines 1 and 4) are approximate.   The neutrino flux in column 
5 is intended above the detection threshold of 1.8~MeV.  The rates in the detectors for lines 1 and 4 are obtained
using a 80\% reactor duty factor, averaged over one year.   The assumptions made on the naval reactors are discussed 
in the text.  The conversion from flux to count rate in the detectors includes the effects of the different 
scintillator composition (100\% pseudocumene for Borexino and 20\% pseudocumene + 80\% paraffin for KamLAND).
$^{\ast}$The distance in this case is calculated to the nearest sea-depth of 30~m.}
\label{tab:backgrounds}
\end{table*}

Naval reactors are mainly installed in submarines, providing underwater endurance and larger power density than
conventional engines. Many aircraft carriers and few other large military vessels are also nuclear powered.
While, in general, nuclear propulsion for civilian vessels has only reached the demonstration stage, a notable
exception is represented by the Russian fleet of large icebreaker that today counts some seven operational ships.
Also in this case the large amount of power and the endurance make nuclear propulsion particularly appropriate
for forcing routes through the northern coast of Siberia.

While many details of the design and operation of naval reactors are not available in the open literature,
for the purpose of this paper it will be sufficient to consider the thermal power and, in some cases, a crude 
estimate of the fuel composition, parameters that are available at least for some reactor types.    Typical 
thermal powers are available for Russian vessels~\cite{Bellona} and can be estimated from the quoted 
``shaft-power''~\cite{warships} for American aircraft carriers assuming a thermodynamical yield of 30\% from
thermal power to mechanical (shaft) power.
In order to estimate the backgrounds from naval reactors to existing experiments we will consider a ``Typhoon'' 
class Russian submarine to transit in the closest
tract of sea to KamLAND (Toyama Bay, Sea of Japan) and Borexino (Adriatic Sea).   In addition we will consider 
the effect on these experiments of a US aircraft carrier entering a large military harbor, such as there exist 
both in Japan (Yokosuka) and Italy (Naples).       ``Typhoon'' subs are the largest submarines ever built.
They are propelled by two reactors with total power of 0.38~GW$_{th}$.   While other subs have lower power
(and often one reactor only), we will use 0.38~GW$_{th}$ as an upper limit for power installed in submarines.    
As far as we know, both the Sea of Japan and the Adriatic are possibly visited by Russian and/or western 
submarines for unknown periods and at unknown times.    

As discussed above we can calculate that the USS Enterprise 
uses $\sim$0.84~GW$_{th}$ for propulsion, generated by two of its eight reactors.   Possibly another reactor
(and $\sim$0.42~GW$_{th}$) is used to produce steam for the catapults launching aircrafts. Other US carriers have
marginally lower power.   At least in principle these ships are easy to detect and their location is not kept 
secret, so we will not examine the case of one of such ships transiting close to shore at the point of closest 
approach to the experiments (however we point out that at present none of the experiments is keeping track of the 
whereabouts of such ships).    To estimate an order of magnitude for the background we will simply assume that 
one of these ships, with only the propulsion power of $\sim$0.84~GW$_{th}$ is in transit at Yokosuka or Naples.    
Although presumably ships do not enter harbors at full power, the number of surface and submarine ships at 
these two locations is probably larger than one at any given time.
A comparison between the signal from fixed reactors and the background from naval reactors is given in 
Table~\ref{tab:backgrounds}.    
For simplicity the flux and count rates from naval reactors have been estimated by assuming the same 
anti-neutrino spectrum produced by stationary power reactors.   Differences in fuel composition, which will
be discussed later, do not alter appreciably the situation.

Keeping in mind that systematics of at least 2-3\% have to be expected in this kind of 
experiments~\cite{Bemporad_etal} it is clear from the table that the effect of nuclear vessels is
not very important at KamLAND, although some method will have to be devised to assign a systematic.      
The situation is already somewhat different for Borexino, in which case a reliable estimate of this 
systematic error appears essential and rather difficult.
The possible presence of naval reactors seems even more problematic for future experiments
that might be built for precision measurements of oscillations.    It is likely that both
Gran Sasso and Kamioka will result unsuitable for such future endeavors.     Continental
location such as Baksan, Heilbron~\cite{schoenert-lasserre}, Homestake, Soudan and WIPP at Carlsbad
are a-priori much more suitable from this point of view.
We also note, in passing, that while naval reactors appear to be a problem for precision measurements
of reactor neutrinos, their effect is not sufficient to allow reliable and efficient detection of the 
vessel carrying them, at least for any conceivable detector size.

We now discuss the possibility of using a naval reactor as a mobile source of anti-neutrinos
in oscillation experiments.    Variable baseline experiments are known to have a number of advantages over fixed
ones.   However, as long baseline experiments become larger and require massive shielding, the possibility of 
moving the detector becomes technically difficult.
Neutrino sources also are considered un-movable, both in the case of accelerators and reactors.     Since the oscillatory
term is proportional to $L/E$, where $L$ is the baseline and $E$ is the neutrino energy, a change in energy
is equivalent to a change in baseline.   Indeed this feature is used in accelerator experiments as well as reactor 
experiments, where $E$ naturally spans the region 1.8 -- 9~MeV.     With some luck the $\Delta m^2$
for $\bar\nu_e - \bar\nu_X$ oscillations will be such that KamLAND will observe a spectrum vastly distorted by
oscillations.   However the range of energies accessible with a reactor or a single accelerator beam-line pointing
in a fixed direction is limited.   In addition, in the case of reactors, the construction of several detectors at 
different distances does not allow one to cancel the very important systematic error contributions from backgrounds
and detector efficiencies.   A naval reactor used as anti-neutrino source with a detector installed underground
near a waterway may come close to the ideal setup.   While it appears out of question to consider using a military 
vessel for this purpose, a ship from the Russian nuclear icebreaker fleet may well be made available, particularly
outside their operating season.

\begin{figure}[htb!!]
\begin{center}
\mbox{\epsfig{file=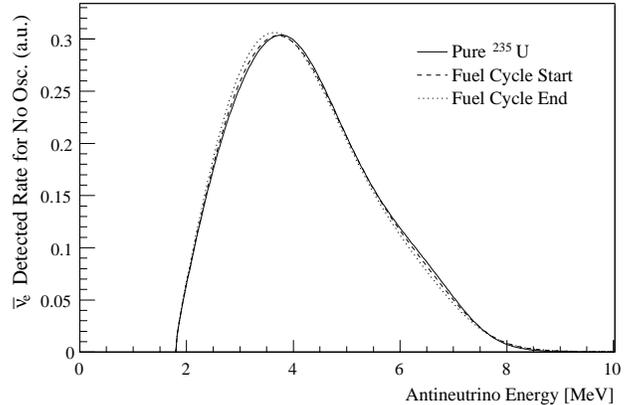,height=6.7cm}}
\vskip 0.1cm
\caption{Energy spectrum of anti-neutrinos produced by different types of reactor fuel.   The two broken lines
correspond to the fuel composition of a commercial power reactor at the beginning and at the end of a fuel
cycle.   The solid line represents the spectrum for a hypothetical core made of  $^{235}$U only.}
\label{fig:reactordist}
\end{center}
\end{figure}

Five ``Arktica'' class nuclear icebreakers are operated commercially by the Murmansk Shipping Company 
(MSCO)~\cite{MSCO}.    
While the main mission of the fleet is to guide cargo vessels along the northern coast of Siberia, some of the ships
have been employed in a variety of tasks, including taking tourists to the North Pole.  Each of these icebreakers
is powered by two 100~MW$_{th}$ nuclear reactors for a total power of 200~MW$_{th}$.   The cores are loaded with uranium
enriched to 20\% in $^{235}$U~\cite{Sivintsev} approximately every three years.      For reference in 
Figure~\ref{fig:reactordist} we compare the anti-neutrino spectrum produced by a hypothetical reactor built 100\% of 
$^{235}$U with the more familiar case of a commercial reactor at the beginning and at the end of a fuel cycle 
(most commercial reactors at the end of a fuel cycle replace 1/4 to 1/3 of their fuel with uranium enriched
to 4\% in $^{235}$U).     The spectrum from the icebreaker's reactors will lay somewhere between those presented 
in the figure depending upon the burn-up stage of the core.    While the exact spectrum will have to be accurately 
known for an actual experiment, for the purpose of this analysis we will simply assume a 100\% $^{235}$U core.

The lower power levels available from naval reactors call for a larger detector.   We will assume to have
a liquid scintillator detector with 20~ktons fiducial mass and a hydrogen to carbon ratio of 2 (as in KamLAND).
Scintillation detectors of this size have been discussed~\cite{svoboda} and would fit in cavities of the size
occupied by SuperKamiokande.     At 2~km distance from the naval reactor such a detector would observe 
$3.9\times 10^5 \bar\nu_{e}\rm yr^{-1}$.   Such a rate, scaling with $1/L^2$, would reduce to 
$1.6\times 10^4 \bar\nu_{e}\rm yr^{-1}$ at 10~km and $9.8\times 10^2 \bar\nu_{e}\rm yr^{-1}$ at 40~km.
The detector would have to be built at an underground location very far from permanent nuclear installations
but accessible by water.    An initial extended period of data with no icebreaker in the vicinity would be 
used to measure the background and other natural phenomena.    We can assume that at the end of this period
the background will be known with negligible error, as it was in the case of Chooz.   The naval reactor
would then be brought to the point of closest proximity, that we assume to be 2~km (closer distances are excluded by the
need for detector shielding).     A 1~month measurement at this distance would give a flux/detector efficiency 
calibration at the 0.5\% level.   Measurements would then be performed at a variety of intermediate distances,
first in short periods to identify the optimal baseline, then, probably, for an extended period at this optimal place.
The calibration run may then be repeated before switching to a second phase in the program, with no reactor, 
when geological neutrinos and astrophysical sources would be addressed without reactor background.    
The large size of the detector would be justified also by this second phase of experimentation.   It may be 
appropriate during the lifetime of the detector to periodically bring the icebreaker in for calibrations with neutrinos.

\begin{figure}[htb!!]
\begin{center}
\mbox{\epsfig{file=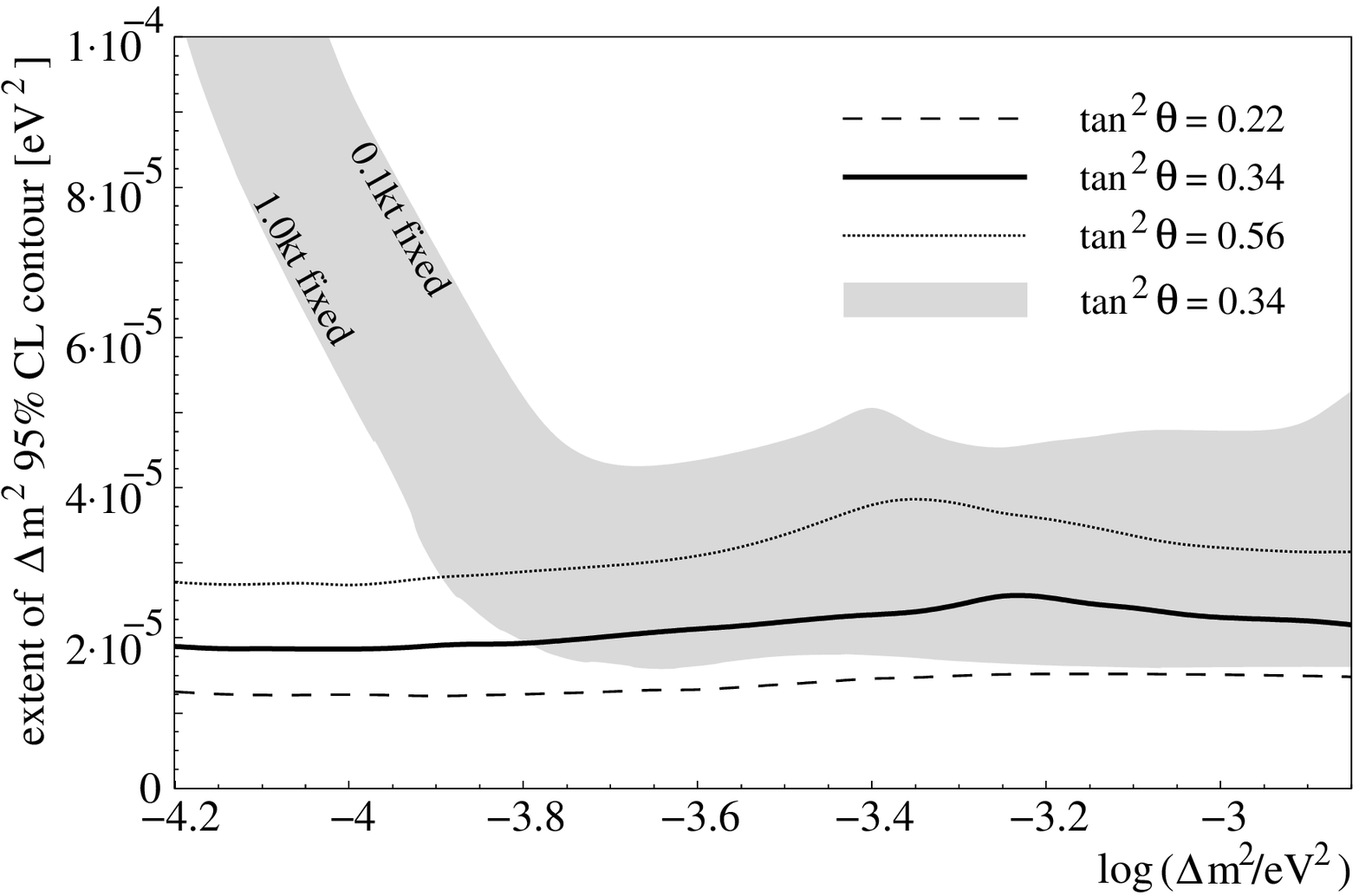,height=5.5cm}}
\mbox{\epsfig{file=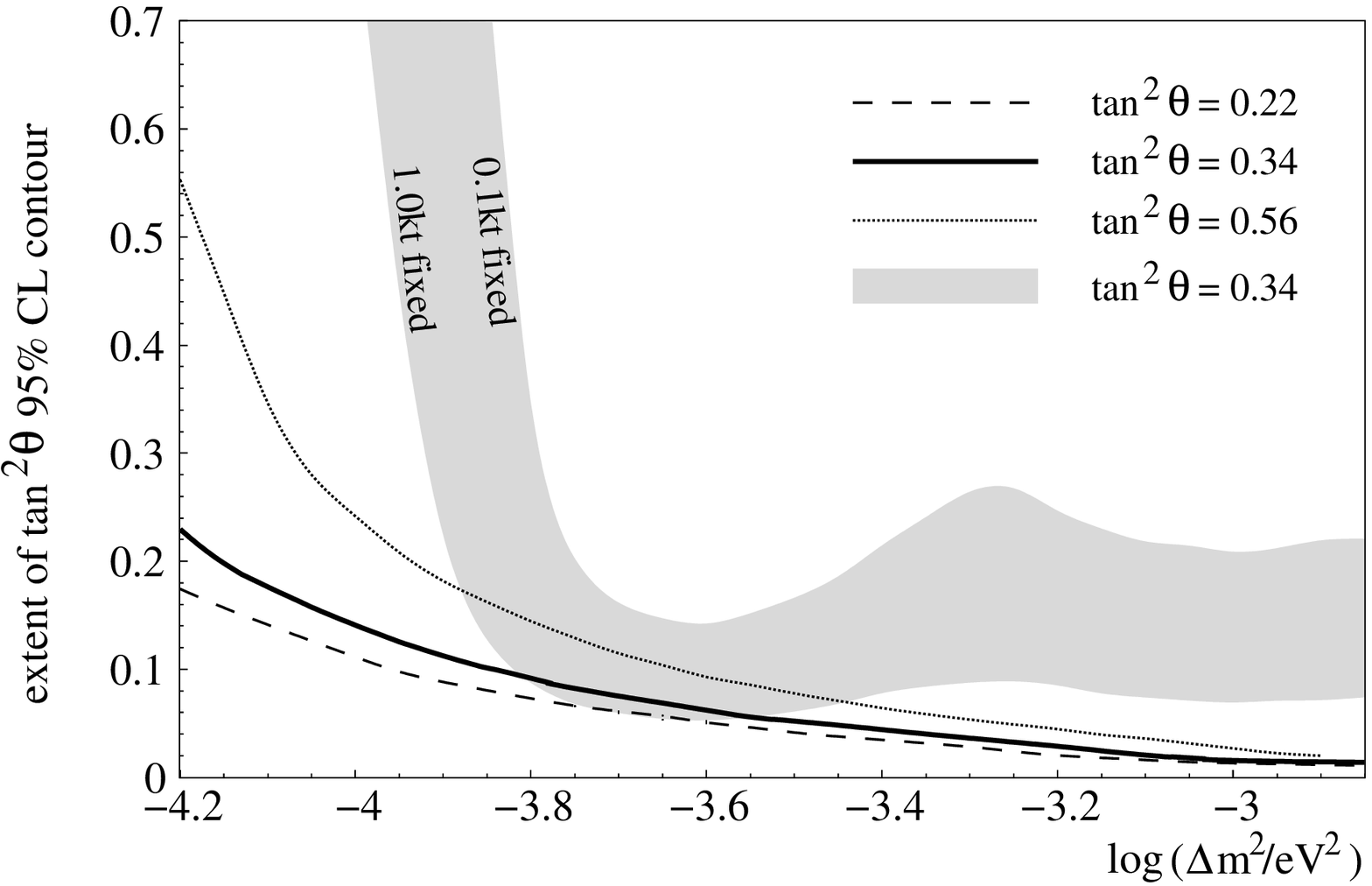,height=5.5cm}}
\vskip 0.1cm
\caption{Estimated 95\% CL errors in $\Delta m^2$ (upper panel) and $\tan^2\theta$ (lower panel) for tunable
baseline experiments compared to the case of a fixed baseline experiment as a function of $\Delta m^2$.
The three curves for the tunable baseline case refer to $\tan^2 \theta = 0.22$, 0.34 and 0.56, 
as consistent with the central value and the extremes of the LMA-MSW solar neutrino solution~\protect{\cite{SNO_daynite}}.   
Only the central value is used for the case of the fixed baseline experiment represented by the shaded region
bounded above and below by, respectively, the case of a 100~ton and 1~kton detector.}    
\label{fig:deltas}
\end{center}
\end{figure}

A detailed running strategy for the reactor neutrino oscillation experiment would have to be established
also taking into account the knowledge of the neutrino mixing matrix at the time of the experiment.
Here we estimate the sensitivity of this scheme assuming that data is collected 
for one year at the optimal baseline for each hypothetical value of $\Delta m^2$.   For each pair of $\Delta m^2$ and 
$\tan^2\theta$ we then estimate the 95\%~CL contours of the measurement.  In Figure~\ref{fig:deltas} we plot 
the projections of such contours (errors in $\Delta m^2$ and $\tan^2\theta$) as functions of $\Delta m^2$ along
with the same quantities for 100~ton~\cite{schoenert-lasserre} and 1~kton detectors at a fixed baseline of 20~km.    
The three curves for the variable baseline case refer to different mixing angle values.     While in the fixed 
reactor case a systematic error of 2.7\% is used, in the case of the movable reactor we assume a 0.5\% systematic 
error (consistent with the superior ability to measure backgrounds and efficiency).    Appropriate statistical 
errors are applied to each case.

Clearly the sensitivity of the fixed baseline case is dominated by the oscillation pattern, while the
variable baseline case becomes limited by statistics at large distances (small  $\Delta m^2$).
It should be noted here that the small 0.5\% systematic error assumed here would be dominated by the 
relative reactor power measurement between calibration and data taking phases.     It appears like 
such relative error for reactors mounted on icebreakers is currently only of about 
10\%~\cite{Sivintsev}.   Therefore some engineering would be needed to install precise temperature and flow
gauges on the cooling water loops.    Commercial nuclear power plants typically measure their power to 
better than 1\% in absolute terms, so the error we quote should be easily attainable.
The possibility of running the reactors at full power with the ship stationary (i.e. without dissipating 
$\sim$1/3 of the power in moving the ship) would also have to be investigated.

Several general locations would have to be investigated in order to find the optimal location for an experiment
of this kind.     In general it appears like the subarctic regions may indeed offer the best chances for
underground sites near closed waterways and surrounded by a geophysics that is optimal for the later study of
geological neutrinos.    Furthermore these regions are far from permanent nuclear power plants.   Ideally
one of the Siberian rivers would ensure that no other nuclear powered vessel approaches the detector, although 
similar conditions could also be satisfied in a number of Canadian bays accessible from the Arctic Ocean.



We would like to thank Mr. A.~Chernov, Mr. G.~Kornilov and Dr. Yu.V.~Sivintsev
for the help in understanding the performance of reactors on Russian icebreakers
and Drs. N.~Sleep and P.~Vogel for critical readings of the manuscript.
This work has been supported, in part, by US DoE Grant DE-FG03-90ER40569-A019.

\end{document}